\documentclass[prd,preprintnumbers,amsmath,amssymb,usenatbib,nofootinbib,superscriptaddress,showkeys,showpacs,11pt]{revtex4-1}

\usepackage{amsmath}
\usepackage{amsfonts,color}
\usepackage{amssymb,float}
\usepackage{graphicx}
\usepackage{subfigure}
\usepackage{appendix}
\usepackage[colorlinks]{hyperref}
\usepackage{xcolor}
\usepackage{orcidlink}
\usepackage{epsf}
\usepackage{bm}
\usepackage{epstopdf}
\usepackage{natbib}
\usepackage{lipsum}
\newcommand{\longeq}{\scalebox{5}[1]{=}}

\begin{document}

\title{Viscous interacting and stability on dark matter Bose-Einstein condensation with modified Chaplygin gas}

\author{E. Mahichi}
\email{elhammahichi@gmail.com}

\author{Alireza Amani\orcidlink{0000-0002-1296-614X}}
\email[Corresponding author:~] {a.r.amani@iauamol.ac.ir}

\author{M. A. Ramzanpour}
\email{m.ramzanpour@iauamol.ac.ir}
\affiliation{Department of Physics, Ayatollah Amoli Branch, Islamic Azad University, Amol, Iran}

\date{\today}


\begin{abstract}
In this paper, the viscous cosmological dynamics are studied in the presence of dark matter Bose-Einstein Condensation (BEC) by curved-FRW background. For this purpose, we use the BEC regime rather than the normal dark matter (the cold dark matter or the barotropic dark matter) with the dark matter Equation of State (EoS) as $p_{dm} \propto \rho_{dm}^2$, which arises from the gravitational form. Therefore, we obtain the corresponding continuity equations with the existence of the universe components by considering an interacting model with modified Chaplygin gas. Afterward, we derive the energy density and the pressure of dark energy in terms of the redshift parameter. And then, by introducing a parametrization function and fitting it with 51 supernova data with the likelihood analysis, we find the cosmological parameters versus redshift parameter. In what follows, we plot the corresponding dynamic graphs proportional to redshift, and then we represent the universe is currently undergoing an accelerated expansion phase. Finally, we explore the stability and the instability of the present model with the sound speed parameter.
\end{abstract}

\maketitle
\section{Introduction}\label{I}
As we know, the contents of the universe are dark matter, dark energy, visible matter, in which visible matter consists of the galaxies, stars, planets and every other visible object. Dark energy is a kind of mysterious energy that supposed to surround all the space and increased the expansion velocity of the universe. So, for describing the observed accelerated expansion of the universe, it is necessary to have a strong negative pressure. This issue was first discovered in a type Ia supernova (SNe Ia) \cite{Riess_1998}, and then continued with other approaches such as more supernovae, cosmic microwave background, and large scale structure \cite{Perlmutter_1999, Bennett_2003, Tegmark_2004}. Many papers have been studied on dark energy by models such as cosmological constant \cite{Weinberg-1989, Sadeghi1-2009, Setare-2009}, quintessence \cite{Battye-2016, Li-2012, Khurshudyan-2015}, phantom \cite{Caldwell-2002, Sadeghi-2013, Sadeghi-2014, JSadeghi-2015}, quintom \cite{Guo-2005, Setare1-2009, Amani-2011}, tachyon \cite{Sen-2002, Sen1-2002, Iorio-2016, Faraoni-2016}, modified gravity \cite{Sadeghi2-2016, Wei-2009, Amani1-2011, Li-2004, Campo-2011, Hu-2015}, holography \cite{Fayaz-2015, Saadat1-2013, KhurshudyanJ-2014, KhurshudyanB-2014, Morais-2017, Zhang1-2017, Bouhmadi1-2016}, new agegraphics \cite{KhurshudyanJ1-2015, SadeghiKhurshudyanJ-2014, SadeghiKhurshudyanJ1-2014, Sadeghi-2009}, bouncing theory \cite{Sadeghi-2010, Amani-2016, Singh-2016}, teleparallel gravity \cite{Sahni-2003, pourbagher1-2020} and braneworld models \cite{Setare-2008, Brito-2015}.

However, dark energy is about three-quarters of the total matter-energy within  the universe, Dark matter is about a quarter, and the rest of it is ordinary matter. But today, another interesting challenge in theoretical physics is understanding the nature of dark matter. All researchers think dark matter is non-baryonic in the universe and is probably discovered by some undiscovered subatomic particles. Dark matter is introduced as an invisible mass or dark because it does not appear to interact with observable electromagnetic radiation such as light, and is thus invisible to the entire electromagnetic spectrum. As a result, its gravitational effect describes the formation and evolution of galaxies and clusters, as well as large-scale structures in the whole universe. Namely, the experimental evidence represents us that dark matter is emerged due to gravitational pull on ordinary matter. Moreover, the formation and evolution of the contents of the universe arise from other observational evidence such as the gravitational lensing and the cosmic microwave background \cite{Kaiser-1993,Massey-2010, Seljak-1999, Arkani-2009}.

The hypothesis of dark matter has proposed to explain the difference between the calculated mass for giant celestial bodies by the two methods of gravity and the luminous matter within them (gas, stars, and dust). Also, this hypothesis ascribes the appearance of cosmology to the Bose-Einstein Condensation (BEC), and it is believed that BEC can be a window for better understanding the universe, dark matter, and dark energy. Hence, BEC is a state of matter in which dilute Bose gas (these are particles that follow Bose-Einstein statistics and do not follow Pauli's exclusion principle) is cooled into very low temperatures ($0.01 ^\circ K$). Because of the low temperature, phase transition happens and a large part of bosons occupy a minimum quantum state, and the macroscopic quantum phenomenon appears at that point. Cooled bosons collapse to each other and superparticles that have microwave behavior emerge. This BEC matter is very fragile and the velocity of light in it is very low. It should be noted that Refs. \cite{Anderson-1995, Davis-1995, Bradley-1995} observed a sharp peak in the velocity distribution of the alkali atoms as vapors of rubidium and sodium below the critical temperature, indicating that condensation had occurred, i.e., the alkali atoms condensed in the same ground state. Therefore, the aforesaid descriptions could connect between BEC phenomena and the cosmic evolution of dark matter. Namely, we suppose that dark matter is content a boson gas below the critical temperature, which they will form a BEC, that exceeds the temperature of the universe at all eras. Therefore, some of the bosons fall to the same ground state in the very early universe, which may be considered as viable candidates for dark matter. In fact, once the critical temperature of the boson gas is less than the critical temperature, BEC can always take place at some moment during the cosmic history of the universe (also, see for more details on the critical temperature to Refs. \cite{Harko-2011, Das-2015}). With these descriptions, several papers studied the various aspects of BEC for the evolution of the universe \cite{Harko-2011, Das-2015, Fukuyama-2008, Li-2014, Boehmer-2007, Das-2018}.

As noted above, BEC introduced as one of the models for describing the source of dark matter, but the nature of dark energy could arise from the various fluids such as perfect fluid, viscous fluid, Chaplygin gas, and so on, that one explains the cosmic evolution and is a candidate as an alternative to the standard cosmology. Therefore, between these fluids, we choose the models of bulk viscosity and Chaplygin gas as an interacting fluid with negative pressure \cite{Fabris-2002, Gorini-2003, Arbab-2003, Sadeghiam_2013}. It is interesting to know that the concept of the Chaplygin gas comes from the Nambu-Goto action in the string theory \cite{Ogawa-2000}. Chaplygin gas model extended to generalized Chaplygin gas \cite{Bertolami-2004,  Marttens-2017, Amaniali-2013, SaadataB-2014}, modified Chaplygin gas \cite{Jawad-2017, Naji-2014, Amani-2013, Rezaei-2017}, and generalized cosmic Chaplygin gas \cite{Chakraborty-2007, Amani-2014}, herein we consider the modified Chaplygin gas model that is the combination of the barotropic model and Chaplygin gas model. The advantage of the modified Chaplygin gas model is that corresponds to observational data and unify the dark matter and dark energy. The components of the universe considered such as dark matter, dark energy, and viscous fluid . As is evident, the modified Chaplygin gas and viscous fluid give a piece of important information about the dark energy that leads to an understanding of the late universe, and also BEC helps to realize the dark matter that leads to an understanding of the early universe. Therefore, as mentioned, the origin of dark matter arises from the concept of BEC, which could lead us to find a good motivation for the present job. Then, in this paper, we intend to study an viscous interacting model between the modified Chaplygin gas, dark matter, and dark energy.

This paper is organized as the following:
In Sec. \ref{II}, we review the general form of the Einstein equation with the cosmological constant in the curved-FRW metric. In Sec. \ref{III}, we consider the interaction between the modified Chaplygin gas, dark matter, and dark energy, followed by we will obtain the energy density and the pressure of dark energy as well as the EoS. In Sec. \ref{IV}, we consider the nature of dark matter as BEC, one yield that the dark matter pressure is proportional to the second-order energy density of dark matter. In Sec. \ref{V}, the corresponding model reconstructs by the redshift parameter, and then determine the cosmological parameters by using the observational data, and thence the stability analysis is investigated. Finally, in Sec. \ref{VI} we will give a summary of our model.

\section{The field equations}\label{II}

We start with the Einstein equation in presence of the cosmological constant $\Lambda$ in the following form
\begin{equation}\label{eineq1}
G_{\mu \nu} + g_{\mu \nu} \Lambda = 8 \pi G T_{\mu \nu},
\end{equation}
where $G_{\mu \nu} = R_{\mu \nu} - \frac{1}{2} g_{\mu \nu} R$ is the Einstein tensor and depends on the geometry of space-time (by metric tensor $g_{\mu \nu}$) with the distribution of its inner matter (by energy-momentum tensor $T_{\mu \nu}$), $R$ and $R_{\mu \nu}$ are the Ricci scalar and the Ricci tensor, respectively, here we consider $ \kappa^2 = 8 \pi G$ and $c=1$. Here, we consider the Friedmann-Robertson-Walker (FRW) curved-universe by the following metric
\begin{equation}\label{ds21}
ds^2 = -dt^2 + a^2(t) \left(\frac{dr^2}{1 - k r^2} + r^2 d\theta^2 + r^2 \sin^2 \theta d\phi^2 \right),
\end{equation}
where $a(t)$ is the scale factor (dimensionless quantity), and constant $k$ indicates the curvature of the space, so $k = 0$, $1$ and $-1$ represent flat, spherical, and hyperbolic space, respectively, which its unit is in terms of $L^{-2}$ in $SI$ units or $M^2$ in Planck units.
Now, we consider a realistic model for the evolution of the universe as viscous fluid or bulk viscosity. We note that the bulk viscosity of fluid gives rise to bring about resistance fluid flow in the universe. Therefore, the bulk viscosity affects the cosmic pressure. In that case, the energy-momentum tensor reads
\begin{equation}\label{tmunu1}
T_{\mu \nu} = (\rho_{tot} + p_{tot} + p_{b}) u_\mu u_\nu + \left(p_{tot} + p_{b}\right)\, g_{\mu \nu},
\end{equation}
where $\rho_{tot}$ and $p_{tot}$  are respectively the total energy density and the total pressure inside the universe, and $p_{b} = -3 \xi H$ represents the pressure of the bulk viscosity in which $\xi$ is a positive constant in terms of $Pa.s$ ($M  L^{-1} T^{ -1}$) in $SI$ units or $M^3$ in Planck units, and $u_\mu$ is the velocity 4-vectors which be written as $u^\mu u_\nu = -1$. In that case, we can write the energy-momentum non-zero components as
\begin{equation}\label{enmom1}
T^0_0 = -\rho_{tot}, \,\,\,\,\, T^i_i = \overline{p}_{tot},
\end{equation}
where $\overline{p}_{tot} = p_{tot} - 3 \xi H$.

By inserting Eqs. \eqref{ds21} and \eqref{enmom1} into Eq. \eqref{eineq1} we can obtain the Friedmann equations in the following form
\begin{subequations}\label{fried1}
\begin{eqnarray}
 &3 H^2 + \frac{3 k}{a^2} -  \Lambda = \kappa^2 \, \rho_{tot},\label{fried1-1}\\
 &3 H^2 + 2 \dot{H} +\frac{k}{a^2} -  \Lambda = -\kappa^2 \, \overline{p}_{tot},\label{fried1-2}
\end{eqnarray}
\end{subequations}
where $H = \frac{\dot{a}}{a}$ is the Hubble's parameter. We can write the total continuity equation in the following form
\begin{equation}\label{conteq1}
{\dot \rho _{tot}} + 3{\mkern 1mu} H{\mkern 1mu} ({\rho _{tot}} + \overline{p}_{tot}) = 0.
\end{equation}
The equation of state (EoS), $\omega_{tot}$ and the deceleration parameter, $q$ are written for the total universe by
\begin{subequations}\label{eosdec1}
\begin{eqnarray}
 &\omega_{tot} = \frac{\overline{p}_{tot}}{\rho_{tot}},\label{eosdec1-1}\\
 &q = -1 - \frac{\dot{H}}{H^2}.\label{eosdec1-2}
\end{eqnarray}
\end{subequations}
In next section, we will explore interacting model between the components of the universe.

\section{Interacting model}\label{III}
In this section, we intend to consider that contents of the universe dominate with the dark matter, the dark energy and the modified Chaplygin gas (MCG) in the following form
\begin{subequations}\label{rotot1}
\begin{eqnarray}
 &\rho_{tot} = \rho_{ch} + \rho_{dm} + \rho_{de},\label{rotot1-1}\\
 &p_{tot} = p_{ch} + p_{dm} + p_{de},\label{rotot1-2}
\end{eqnarray}
\end{subequations}
where indices of $ch$, $dm$ and $de$ are the modified Chaplygin gas, the dark matter and the dark energy, respectively. It should be noted that the universe dominates from an exotic matter called the MCG with the following EoS
\begin{equation}\label{mcg1}
p_{ch} = A \rho_{ch} - \frac{B}{\rho_{ch}^n},
\end{equation}
 where $A$ and $B$ are positive constants, and $n$ is in the range of $0 < n \leq 1$.
As we mentioned, in a realistic universe there is an interaction between components of the universe, so this interaction gives rise to appear an energy flow between components. Nevertheless, we can separately write the continuity equations for the universe components in the following form
\begin{subequations}\label{contin2}
\begin{eqnarray}
 &\dot{\rho}_{ch}+3 H (\rho_{ch} + p_{ch}) = Q_1,\label{contin2-1}\\
 &\dot{\rho}_{dm}+3 H (\rho_{dm} + p_{dm}) = Q_2 - Q_1,\label{contin2-2}\\
  &\dot{\rho}_{de}+3 H (\rho_{de} + p_{de} - 3 \xi H) = -Q_2,\label{contin2-3}
\end{eqnarray}
\end{subequations}
where $Q_1$ and $Q_2$ are interaction terms called the energy flow between the universe components which is represented as a phenomenological descriptor. We should consider that the interaction terms in the view of dimensional are proportional to the multiplication between the Hubble parameter and density energy. Because dark matter attracts, dark energy repels, and also dark matter pulls matter inward, dark energy pushes it outward, thus the energy flow transfers from dark matter to dark energy. So in this job, we assume that the interaction terms as $Q_1 = Q_2 = 3 b^2 H \rho_{ch}$. This choice tells us that the energy flow transfers between components of dark energy and matter, and the component of dark matter has no interaction with other components. In that case, the MCG energy density obtains by substituting Eq. \eqref{mcg1} into \eqref{contin2-1} as
\begin{equation}\label{rhoch1}
\rho_{ch} = \rho_{ch}^0 \left[ \frac{\frac{B}{\eta} + \alpha a^{-3 \eta (n+1)}}{\frac{B}{\eta} + \alpha a_0^{-3 \eta (n+1)}}\right]^{\frac{1}{n+1}},
\end{equation}
where $\rho_{ch}^0 = \left[\frac{B}{\eta} + \alpha a_0^{-3 \eta (n+1)}\right]^{\frac{1}{n+1}}$, $\eta = 1 + b^2 + A$, and $a_0$ and $\alpha$ are the current scale factor and a constant integral, respectively.

In what follows, we obtain the dark energy component from the Eqs. \eqref{fried1} and \eqref{rotot1} in the following form
\begin{subequations}\label{rhop2}
\begin{eqnarray}\label{rhop11}
 &\rho_{de} = \frac{1}{\kappa^2}\left(3 H^2 + \frac{3 k}{a^2} -  \Lambda\right) - \rho_{ch} - \rho_{dm},\label{rhop11-1}\\
 &p_{de} =  -\frac{1}{\kappa^2}\left(3 H^2 + 2 \dot{H} +\frac{k}{a^2} -  \Lambda\right) - p_{ch} - p_{dm} + 3 \xi H,\label{rhop11-2}
\end{eqnarray}
\end{subequations}
where the EoS of dark energy yields
\begin{equation}\label{omegade1}
\omega_{de} = \frac{p_{de}}{\rho_{de}},
\end{equation}
where we clearly see that the EoS of dark energy depends on the Hubble parameter, space-time geometry, the cosmological constant, the bulk viscosity, the energy densities and the pressures of dark matter and MCG. In the next section, we will explore the dark matter as a Bose-Einstein condensate.

\section{The dark matter Bose-Einstein condensation}\label{IV}

In this section, we intend to describe dark matter as one of the universe components from the perspectives of the normal dark matter and the Bose-Einstein Condensation (BEC) dark matter. Therefore, in order to solve the corresponding Friedmann equations, we need to introduce the energy density and the pressure of the dark matter by approaches of the normal dark matter and the BEC dark matter.

As the first approach, we take the normal dark matter in the early universe that consisted of bosonic particles with a mass of $m_{dm}$ and a temperature of $T$  \cite{Harko-2011, Hoga_2000, Madsen_2001}. To use the  Bose-Einstein distribution function, $f(p)={[\exp[(E-\mu)]-1]}^{-1}$, energy density and pressure of non-relativistic bosonic particles  obtained by
\begin{eqnarray}\label{mcg2}
&\rho_{dm} =m_{dm} n_{dm},\\
&p_{dm}=\frac{g}{3 (2 \pi)^3} \int \frac{p^2}{E} f(p) d^3p \approx \frac{g}{6 \pi^2} \int \frac{{p^4}dp}{m_{dm}},
\end{eqnarray}
where $n_{dm} = \frac{g}{(2 \pi)^3} \int4\pi f(p){ p^2}dp$ is the
spatial number density of the particles in kinetic equilibrium, in which $E = \sqrt{p^2 + m^2_{dm}}$, $\mu$, $h$, $g$, $p \approx m_{dm} v_{dm}$,  are the energy, the chemical potential, Planks constant, the number of helicity state, and the momentum of the particles, respectively. Here for simplicity we use the Planck units as $c = \hbar = K_B = 1$. In that case, we can write the equation of state for the normal dark matter in the following form
\begin{equation}\label{mcg3}
p_{dm}= \frac{1}{3} \left<v_{dm}^2\right> \rho_{dm},
\end{equation}
where $\left<v_{dm}^2\right>$ is the average squared velocity of the particles. Notice that Eq. \eqref{mcg3} is the same as a barotropic equation for the dark matter as $\omega_{dm} = \frac{p_{dm}}{\rho_{dm}}$, instead of considering the usual dust case or cold dark matter, i.e, $p_{dm} = 0$ \cite{Mueller_2005}. Thus, the barotropic EoS of the dark matter is related to the average squared velocity of the particles.

Now we consider the BEC dark matter as the second approach. BEC is a state that formed at very low temperatures when a dilute Bose gas at low densities is cooled, and one condenses to the same ground state of a quantum system. While we cool the normal bosonic dark matter below the condensed critical temperature and there are two phases simultaneously. In this case, the laws of quantum mechanics replace the laws of classical mechanics. So that instead of particle behavior, we will have wave behavior. When all the dark matter is transferred to the condensed state, the transition ends, this means that one is in a transition phase or transition of BEC. In other words, all particles occupy a single-particle state and generally transfer from a normal state to a condensed state, with the constant temperature and pressure. Therefore, aforesaid descriptions can become the same as our universe.

Therefore, the physical properties of a BEC described by the generalized Gross-Pitaevskii equation that given by the Refs. \cite{Pitaevskii_2003, Pethick_2008}. After that, Harko \cite{Harko-2011} extended BEC in the gravitational form and could obtain the equation of state for the BEC dark matter as follows:
\begin{equation}\label{mcg4}
p_{dm}= u_0 \rho_{dm}^2,
\end{equation}
where $u_o = 2 \pi l_a /m_{dm}^3$, in which $l_a$ is the s-wave scattering length. Now by inserting Eq. \eqref{mcg4} into Eq. \eqref{contin2-2}, we can obtain the variation of the dark matter energy density in terms of time evolution in the following form
\begin{equation}\label{dmed1}
\rho_{dm}= \frac{c_0}{ a^3 - c_0 u_0},
\end{equation}
where $c_0$ is an integral constant. We can write the aforesaid relationship with the current energy density of dark matter $\rho_{dm}^0$ by
\begin{equation}\label{dmed2}
\rho_{dm}= \rho_{dm}^0 \frac{a_0^3 - c_0 u_0}{ a^3 - c_0 u_0},
\end{equation}
where $\rho_{dm}^0 = \frac{c_0}{ a_0^3 - c_0 u_0}$, and $a_0$ is the present scale factor.


\section{Reconstructing and observational constraints}\label{V}

In this section, we are going to solve the Friedmann equations by contributing to the components of the Chaplygin gas, dark energy, and dark matter within the universe. In that case, by putting Eqs. $\eqref{rhoch1}$ and $\eqref{dmed1}$ into the Eq. $\eqref{rhop11-1}$, we can obtain the energy density of the dark energy as follow:
\begin{equation} \label{rhode2}
\rho_{de} =\frac{3 H^2}{\kappa^2} + \frac{3 k}{\kappa^2 a^2} - \frac{\Lambda}{\kappa^2} - \left[\frac{B}{\eta} + \alpha a^ {-3 \eta (n+1)} \right]^{\frac{1}{n+1}} - \frac{c_0}{a^3-c_0 u_0}.
\end{equation}

Now we reconstruct the Friedmann equation by the redshift parameter. For this purpose, we write the relationship between the scale factor and the redshift parameter as $\frac{a_{0}}{a} =1+z$ in which $a_0$ is the current scale factor, so that one immediately get as $\frac{d}{dt} = -H (1+z) \frac{d}{dz}$. Next we introduce a dimensionless parameter for expansion rate in the form
\begin{equation} \label{EHH01}
E(z)=\frac{H(z)^2}{H_{0}^2},
\end{equation}
where $H_0$ is the present Hubble parameter, and the Hubble parameter derivative becomes
\begin{equation} \label{hdot1}
\dot{H} = -\frac{1}{2} H_0^2 (1+z) E',
\end{equation}
where $E' = \frac{dE}{dz}$. The relationships $\rho_{ch}$ and $\rho_{dm}$ versus redshift becomes
\begin{subequations}\label{rhop2}
\begin{eqnarray}
 &\rho_{ch}(z) = \left[
\frac{B}{\eta} + \alpha \left(\frac{1+z}{a_{0}}\right)^ {3 \eta (n+1)}\right]^{\frac{1}{n+1}},\label{rhop2-1}\\
 &\rho_{dm}(z) = \frac{c_{0} (1+z)^3}{a_{0}^3 - c_{0} u_{0} (1+z)^3}.\label{rhop2-2}
\end{eqnarray}
\end{subequations}

The reconstructed Eq. \eqref{rhode2} is rewritten by redshift parameter as
\begin{equation} \label{rhode3}
\rho_{de} =\frac{3 H_{0}^2\, E(z)}{\kappa^2} + \frac{3 k\, (1+z)^2}{\kappa^2\, a_{0}^2} - \frac{\Lambda}{\kappa^2} - \rho_{ch}(z) - \rho_{dm}(z).
\end{equation}

The reconstructed dark energy pressure obtains by substituting Eqs. \eqref{mcg1} and \eqref{mcg4} into Eq. \eqref{rhop11-2} as follow:
\begin{eqnarray} \label{pede1}
&p_{de} = -\frac{3 H_{0}^2\, E(z)}{\kappa^2} + \frac{H_0^2\, (1+z)\, E'}{\kappa^2} + \frac{\Lambda}{\kappa^2} - \frac{k\, (1+z)^2}{\kappa^2 a_{0}^2} - A \rho_{ch}(z) + \frac{B}{\rho_{ch}^n(z)} - u_0\, \rho_{dm}^2(z) + 3 \xi H.
\end{eqnarray}

Hereinafter we proceed by flat-background, then the EoS of dark energy in accordance with Eq. \eqref{omegade1} yields
\begin{equation} \label{omegade2}
\omega_{de} = -1 - \frac{ \kappa^{-2} H_0^2\, (1+z) E'(z) + (A+1)\, \rho_{ch}(z) - B\, \rho_{ch}^{-n}(z) - u_0\, \rho_{dm}^2(z) + 3 \xi H}{3 \kappa^{-2} H_0^2\, E(z) - \kappa^{-2} \Lambda - \rho_{ch}(z) - \rho_{dm}(z)}.
\end{equation}

\begin{figure}[h]
\begin{center}
{\includegraphics[scale=.3]{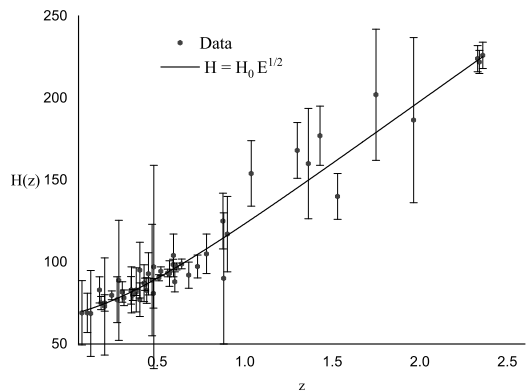}}
\caption{The Hubble parameter in terms of the redshift data (diamond + error bar) and the current model (line).}\label{fig1}
\end{center}
\end{figure}

Now in order to calculate the cosmological parameters, we consider the parametrization function $E(z)$ as a function of the third order polynomial $z$ in the form
\begin{equation} \label{paramet1}
E(z) = A_3 (1+z)^3 + A_2 (1+z)^2 +A_1 (1+z) +A_0,
\end{equation}
where $A_3$ to $A_0$ are determined by using the observational data and have dimensionless units. In this job, coefficients $A_3$ to $A_0$ are only dependent on  the $H(z)$ dataset and them calculate to fit with observational data. Then, insert the obtained parametrization function $ E (z) $ into the Friedmann equations and the corresponding equations to obtain the cosmological parameters. Note that the corresponding parametrization function first proposed by Ref. \cite{Sahni_2003}, after that some papers used in this way \cite{Alam-2007, Setare1-2009}. Next, Ref. \cite{Pourbagher_2019} enhanced it with more supernova data and re-measured coefficients. Now we assign the function $E(z)$ with 51 supernova data that these data collected the Refs. \cite{Blake_2012, Font_2014, Delubac_2015, Alam_2016, Moresco_2016, Farooq_2017, Pacif_2017, Magana_2018}. We note that the $H(z)$ dataset are measured by techniques of galaxy differential age or cosmic chronometer and radial BAO size methods. In what follows, these data set the Hubble parameter in terms of the redshift parameter by interval $0.07 \leq z \leq 2.36$. In order to acquire the parameters $A_3$ to $A_0$, we have to fit the parametrization function $E(z)$ with these data as an experimental constraint. For this purpose, we use the maximum likelihood analysis by introducing the least squares called chi-square value $\chi_{min}^2$ as the simplest shape for the corresponding data analysis. Then, the chi-square value is defined
\begin{equation} \label{likelihood1}
\chi_H^2 = \sum_{i = 1}^{51} \frac{\left(H_{obs}(z_i) - H_{th} (z_i, H_0)\right)^2}{\sigma_H^2 (z_i)},
\end{equation}
where the index $H$ indicates the Hubble dataset, $H_{obs}$ and $H_{th}$ represent the observed value and theoretical of the Hubble parameter, $\sigma_H$ demonstrates the standard error in the observed values, and $H_0 = 68 \,km s^{-1} Mpc^{-1}$ is introduced as the present Hubble constant. Therefore, the parametrization coefficients find with the standard error as $A_3 = -0.07 \pm 0.24$, $A_2 = 1.87 \pm 1.39$, $A_1 = -2.83 \pm 2.34$ and $A_0 = 2.03 \pm 3.97$. Fig. \ref{fig1} shows us the best fitting curve of the current model with the Hubble parameter dataset.

\begin{figure}[h]
\begin{center}\label{AB}
{\includegraphics[scale=.3]{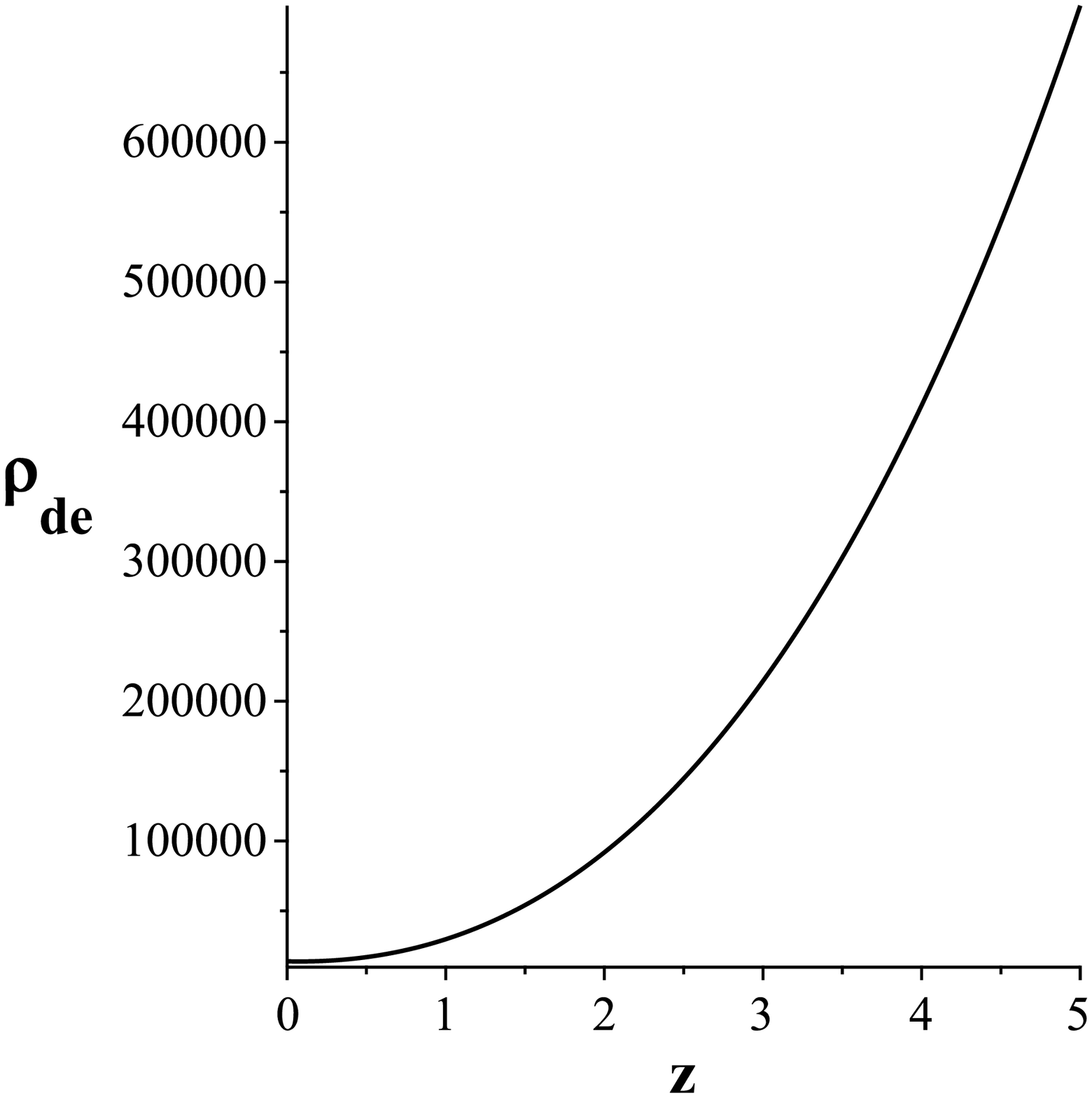}}
{\includegraphics[scale=.3]{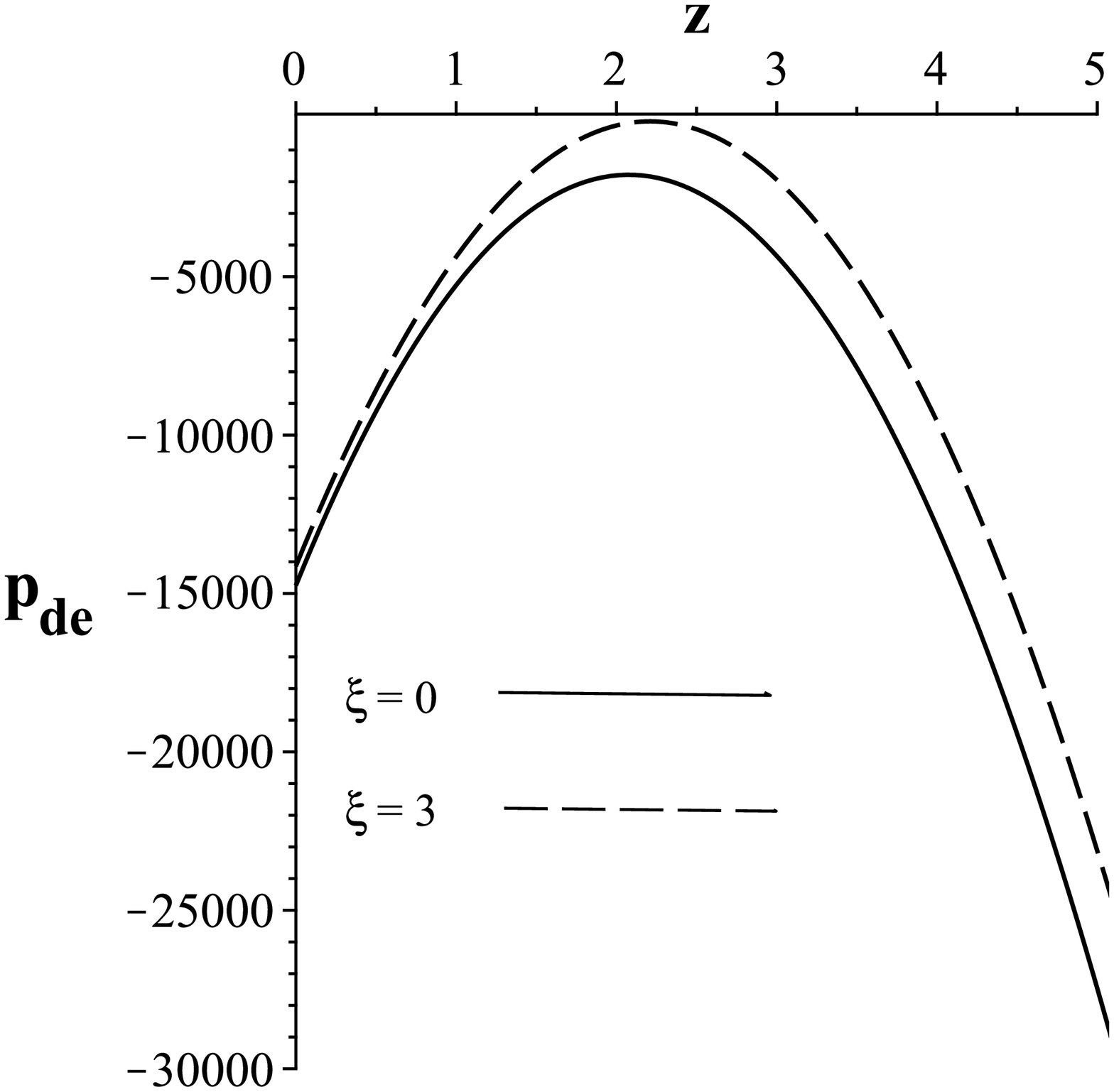}}
\caption{The energy density ($M L^{-1} T^{-2}$ in $SI$ units or $M^4$ in Planck units) and pressure ($M L^{-1} T^{-2}$ in $SI$ units or $M^4$ in Planck units) of dark energy in terms of redshift parameter for $A_3 = 0.17$, $A_2 = 0.65$, $A_1 = -2$, $\alpha = 5$, $a_0 = 10$, $A = 0.5$, $B = 1$, $b = 0.5$, $c_0 = 0.5$, $u_0 = 1$ and $n = 0.75$.}\label{fig2}
\end{center}
\end{figure}

Now, in order to analyze the cosmological parameters, we insert Eq. \eqref{paramet1} into Eqs. \eqref{rhode3}, \eqref{pede1}, and \eqref{omegade2}, which use the corresponding graphs due to the large mathematical relationships. It should be noted that the free parameters have a very important role in the dynamics of the cosmological parameter. Therefore, despite the sensitivity of the issue, the choice of these parameters is that the universe is undergoing an accelerated expansion phase. Moreover, it must satisfy some other constraints such as the  energy density greater than zero and the pressure less than zero in the present epoch, and also the measurement of the EoS crosses from the $-1$ in the late time. In this case, the free parameters are found as $A_3 = 0.17$, $A_2 = 0.65$, $A_1 = -2$, $\alpha = 5$, $a_0 = 10$, $A = 0.5$, $B = 1$, $b = 0.5$, $c_0 = 0.5$, $u_0 = 1$, $n = 0.75$, and $\xi = 0, 3$. It should be noted that parameter $u_0$ is $M^{-3} L$ in $SI$ units or $M^{-4}$ in Planck units, parameter $\xi$ is $Pa.s$ ($M L^{-1} T^{-1}$) in $SI$ units or $M^{3}$ in Planck units, and other free parameters are dimensionless. These free parameters show us that the current model are directly depend on coefficients of Chaplygin gas, dark matter, the interacting term, the cosmological constant and the parametrization function. For this purpose, we plot the graph of Eqs. \eqref{rhode3} and \eqref{pede1} namely respectively the energy density and the pressure of dark energy in terms of the redshift parameter as shown in Fig. \ref{fig2}.

We can see in Figs. \ref{fig2}, the measure of the energy density and the pressure of dark energy are bigger than zero and smaller than zero for the present universe (note that the present universe occurs in $z=0$), respectively. Therefore, the corresponding both of the figures are depending on free parameters that arise from the physical meaning of  Chaplygin gas, dark matter, the interacting model, the cosmological constant, and the coefficients of parametrization function.
\begin{figure}[h]
\begin{center}
{\includegraphics[scale=.3]{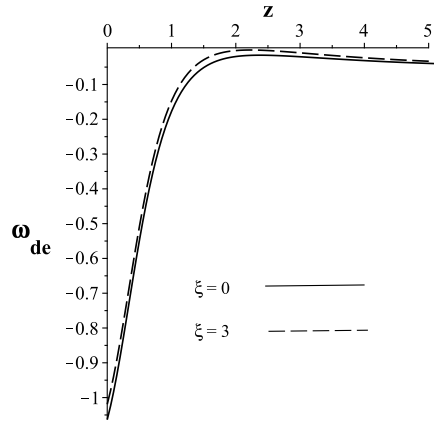}}
\caption{The EoS (dimensionless quantity) of dark energy in terms of redshift parameter for $A_3 = 0.17$, $A_2 = 0.65$, $A_1 = -2$, $\alpha = 5$, $a_0 = 10$, $A = 0.5$, $B = 1$, $b = 0.5$, $c_0 = 0.5$, $u_0 = 1$ and $n = 0.75$.}\label{fig3}
\end{center}
\end{figure}

By using Eq. \eqref{omegade2}, we can plot the dark energy EoS with respect to the redshift parameter as shown in Fig. \ref{fig3}. It should be noted that one of the most important cosmological parameters is introduced  as the EoS parameter that has a main role in various epochs of the universe. This means that whenever $\omega > -1$, $\omega = -1$ and $\omega < -1$ represent  the universe evolution as matter epoch, cosmological constant, and phantom epoch, respectively in which dark energy can be a very strange form of phantom energy. Hence, Fig. \ref{fig3} displays us that the measure of the EoS is $-1.06$ in absence of bulk viscosity and  is $-1.02$  with the existence of viscous fluid for the current time, namely, the universe begins from the quintessence era, and then by crossing the phantom separator line ends to phantom era. Therefore, this result indicates that the universe is undergoing an accelerated expansion phase, which is compatible with the obtained results in Refs. \cite{Amanullah_2010, Scolnic_2018} for EoS.
\begin{figure}[h]
\begin{center}
{\includegraphics[scale=.3]{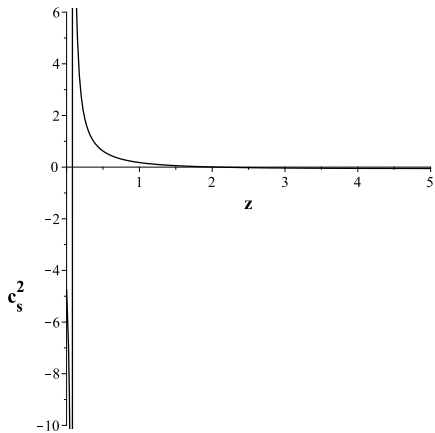}}
\caption{The sound speed parameter (dimensionless quantity) in terms of redshift parameter for $A_3 = 0.17$, $A_2 = 0.65$, $A_1 = -2$, $\alpha = 5$, $a_0 = 10$, $A = 0.5$, $B = 1$, $b = 0.5$, $c_0 = 0.5$, $u_0 = 1$ and $n = 0.75$.}\label{fig4}
\end{center}
\end{figure}

Now, we intend to study the present model from the thermodynamic perspective. For this purpose, we need to examine the stability and the instability of our universe which is considered as a thermodynamics system. We note that stability and instability of a gravitational fluid due to small perturbations lead to growing controlled and non-controlled of energy density in the various universe epochs. In what follows, this thermodynamic system is taken as an adiabatic process, i.e., there is no transferring of heat or mass between within the universe and its outside. In this case, the entropy perturbation is equal to zero, $\delta S = 0$, this means that the perturbed pressure vary only on the energy density in the form
\begin{equation}\label{perturpde}
\delta p_{de}(\rho_{de}, S) = \left. \frac{\partial p_{de}}{\partial \rho_{de}}\right\rvert_{S=const} \delta \rho_{de} + \left. \frac{\partial p_{de}}{\partial S}\right\rvert_{\rho_{de}=const} \delta S \overset{Adi. \,Proc.}{\longeq} c_s^2\, \delta \rho_{de},
\end{equation}
where $c_s^2 = \frac{\partial p_{de}}{\partial \rho_{de}} = \frac{\partial_z p_{de}}{\partial_z \rho_{de}}$ is introduced as the sound speed parameter in a fluid. Then, this parameter is  useful function to determine the stability and instability of the universe with conditions $c_s^2 > 0$ and $c_s^2 < 0$, respectively. By using the corresponding parameter, as shown in Fig. \ref{fig4}, we can plot the variety of $c_s^2$ in terms of the redshift parameter. Fig. \ref{fig4} displays us that the universe is instability in the large redshifts, i.e., in the early epoch. Then, by decreasing of the redshift value, the sound speed parameter becomes more than zero that demonstrates the universe is in a stability phase. Until the late time, the universe is in an instability phase that represent the energy density leads to non-controlled growing. Therefore, we can see in the current model, there is an instability condition in early and late eras but there is stability condition before the last universe. We note that the variety of $c_s^2$  in the two states of viscosity ($\xi = 3$) and non-viscosity ($\xi = 0$) are not very different from each other.

\section{Conclusion}\label{VI}
In this paper, we studied the dark matter Bose-Einstein Condensation by curved-FRW background, and implemented a viscous interacting model between the BEC and modified Chaplygin gas. The Einstein equation was written in the form of the standard cosmology in the presence of the cosmological constant in a bulk viscosity. And then, the corresponding Friedmann equations were obtained in terms of the total energy density and the total pressure. After that, we considered the universe dominated by dark matter, modified Chaplygin gas, and dark energy. Next, the continuity equations separately were written  for the universe components in the presence of the interacting term which one was considered $Q_1 = Q_2 = 3 b^2 H \rho_{ch}$, namely, the energy flow transfers between components of the dark energy and Chaplygin gas, in which the component of dark matter has no interaction with the other components. Nevertheless, we acquired the energy density and the pressure of dark energy in terms of the other components of the universe, the Hubble parameter, and the cosmological constant in the curved-FRW background.

As we know, BEC is a state of matter in which dilute Bose gas is cooled into very low temperatures. This issue gives rise to arise a connection between BEC and dark matter. We took the BEC dark matter rather than the normal dark matter. For this purpose, we considered two approaches for the description of the dark matter as (a) the dark matter pressure has a linear relationship with the energy density as $p_{dm} \propto \rho_{dm}$, (b) the dark matter is proportional to the second-degree energy density as $p_{dm} \propto \rho_{dm}^2$. Herein, we used the second approach by the motivation that one arises in the gravitational form, so this choice is different from the approaches of dust case (cold dark matter) and the barotropic dark matter. In this case, the energy density and the pressure of dark energy reacquired in terms of the corresponding expressions, and then constructed with respect to the redshift parameter.

In what follows, in order to solve the corresponding relationships, we used the observational constraints with 51 supernova data that made us fitted these data to the parametrization function $E(z) = H^2 / H_0^2$ by the likelihood analysis. Then, we plotted the figures of the cosmological parameters versus the redshift parameter depending on the universe components. These figures showed us that the universe is undergoing accelerated expansion and is consistent with observational data. Finally, we analysed our model in the view of the stability and the instability, that the corresponding results represented the universe is in a instability phase for the present time, i.e., the energy density of dark energy is non-controlled in late universe. As a future work, we propose the investigation of the aforesaid model with the modified gravity.


\end{document}